\documentclass[a4paper]{llncs}

\usepackage{graphicx}
\usepackage[table]{xcolor}
\usepackage{subfig}

\begin{document}
\title{IMC: A Classification of Identity Management Approaches} 

\titlerunning{IMC Classification of Identity Management}

\author{Daniela Pöhn\inst{1} 
 \and Wolfgang Hommel\inst{1}
}
\authorrunning{Pöhn and Hommel}
\institute{Bundeswehr Universität München, Research Institute CODE, Munich, Germany
\email{\{daniela.poehn,wolfgang.hommel\}@unibw.de}}

\maketitle

\begin{abstract}
This paper presents a comprehensive classification of identity management approaches. The classification makes use of three axes: topology, type of user, and type of environment. The analysis of existing approaches using the resulting identity management cube (IMC) highlights the trade-off between user control and trust in attributes. A comparative analysis of IMC and established models identifies missing links between the approaches. The IMC is extended by a morphology of identity management, describing characteristics of cooperation. The morphology is then mapped to the life cycle of users and identity management in a further step. These classifications are practically underlined with current approaches. Both methods combined provide a comprehensive characterization of identity management approaches. The methods help to choose suited approaches and implement needed tools.
\end{abstract}

\section{Introduction}\label{introduction}

Thousands of web applications around the world provide different services via the internet. These services require the user to present an identity for authentication, otherwise the user is not able to access them. To manage different users with their identities, identity management (IdM) was introduced as a paradigm more than two decades ago. It focuses on managing usernames, which are used as identifier assigned to users, some sort of credential, usually a password, and further information, like email address and postal address, called user attributes.

Different evolving requirements led to the creation of different models of and protocols for identity management systems (IdMS). While stand-alone organizations run a \textit{centralized} Identity \& Access Management (I\&AM) system, many organizations with collaboration, especially in academia, introduced \textit{Federated Identity Management} (FIM). FIM is an arrangement between multiple entities in order to let users use the same identification data as in their home organization. By FIM, users obtain access to the services provided by partners, called service providers (SPs), within organizational trust boundaries called federations. The often-used Security Assertion Markup Language (SAML) \cite{samloverview} is rather static, whereas OAuth and OpenID Connect (OIDC) \cite{openidconnect} provide a dynamic approach, known for example from Google. Limitations of FIM led to different approaches, like inter-federations (e.g. eduGAIN \cite{edugainstatus}), the use of the Domain Name System (DNS) for discovery and trust, e.g., LIGHTest~\cite{mci/Rossnagel2017}, different assurance frameworks and components. In parallel, \textit{user-centric solutions} were developed. User Managed Access (UMA)~\cite{uma}, an OAuth-based standard, enables the user to control the authorization of data sharing and other protected resources.  The user of \textit{Self-Sovereign Identities} (SSIs) is the ultimate owner of the identity. SSIs are typically realized by decentralized networks, like distributed ledger technologies (DLTs)~\cite{8776589}. Decentralized Identifiers (DIDs)~\cite{w3c} often make use of DLTs.

IdM is one crucial pillar of security frameworks. Several different models and approaches are currently developed and run. Not all approaches fit into one single model, making a categorization challenging. This paper contributes the following improvements: The developed identity management cube (IMC) categorizes different IdM approaches. The cube is broadened by a morphology describing aspects of collaboration within the life cycle. Both categorizations are applied to different protocols and applications. This helps to identify fitting approaches and missing tools for interoperability. It also provides an overview of important aspects during the life cycle, helping stakeholders.

This paper is organized as follows. We discuss related work in Section~\ref{sec:sota}. In Section~\ref{sec:models}, we present a new categorization of IdM and provide a brief classification of current approaches. Additionally, we present a morphology in Section~\ref{sec:addon}, which is then mapped to the life cycle of identities and identity management. The newly developed IMC and the morphology are applied to current approaches and then discussed in Section~\ref{sec:five}. The paper is concluded in Section~\ref{sec:summary} by a summary of the results achieved so far and an outlook to ongoing work.

\section{Related Work}\label{sec:sota}

Yuan Cao and Lin Yang~\cite{5689468} identify three core components for IdM: user, service provider (SP), and identity provider (IdP). The authors further describe the three models isolated, centralized, and federated. According to them, the IdM paradigms can be classified into network-centric paradigm, service-centric paradigm, and user-centric paradigm. Sovrin~\cite{sovrin} sees SSI as next step after isolated, centralized, federated, and user-centric IdM models. In other papers, either the models isolated, centralized, federated, and user-centric or centralized, federated, and decentralized are used.

Boujezza et al.~\cite{7507266} describe a taxonomy for Internet of Things (IoT) by adapting the paradigms and requirements. The authors classify user model, service provider model, and hybrid model, combining user and SP, and further submodels. In contrast, Pal et al.~\cite{8603595} relate IoT identities to things-centric identities. Gao et al.~\cite{7423325} describe an IdM model for big data based on authorization, authentication, identification, and audit modules. Habiba et al.~\cite{Habiba2014} use the IdM requirements taxonomy to classify cloud IdMS. Further approaches have been developed, leading to different directions, which we integrate into our model.

\section{Identity Management Models}\label{sec:models}

The main functionalities of IdM are identification, authentication, and authorization. In most cases, a password is provided for authentication, which fulfils a required complexity or entropy. Second factor, multi-factor, and anonymous are also possible. The authorization is based on policies, which describe whether the user is allowed to access a certain functionality or data. With collaborations, the information about the user is stored at the IdP. The user wants to access a service of the entity SP. Minor entities are trusted third parties (TTPs), attribute authorities (AAs), having additional information about the user, and federation operators, if IdPs and SPs form trust boundaries. As new requirements are evolving, different approaches for IdM have been developed and will be emerging in the future. The existing IdM models do not work for several use cases. Therefore, new models are developed and applied in the following.

\subsection{Analysis of Identity Management Models}

In order to distinguish different IdM approaches, models have been established. These models were updated for user-centric models and partly for SSI. As described in Section~\ref{sec:sota}, the following IdM models are mostly used.

\begin{description}
\item[Isolated:] I\&AM per service.
\item[Centralized:] Network-centric. I\&AM per entity, e.g., with single sign-on (SSO).
\item[Federated:] Application-centric. I\&AM per federation, which is a set of IdPs and SPs. Possible protocols are, e.g., OIDC and SAML.
\item[Decentralized:] User-centric. I\&AM, where the user is in control. Used for FIM in many cases. Possible approaches are UMA and SSI. Decentralized is partly divided into user-centric and SSI.
\end{description}

\begin{figure}
\subfloat[Evolution of Identity Management Models]{
\includegraphics[width=.45\textwidth]{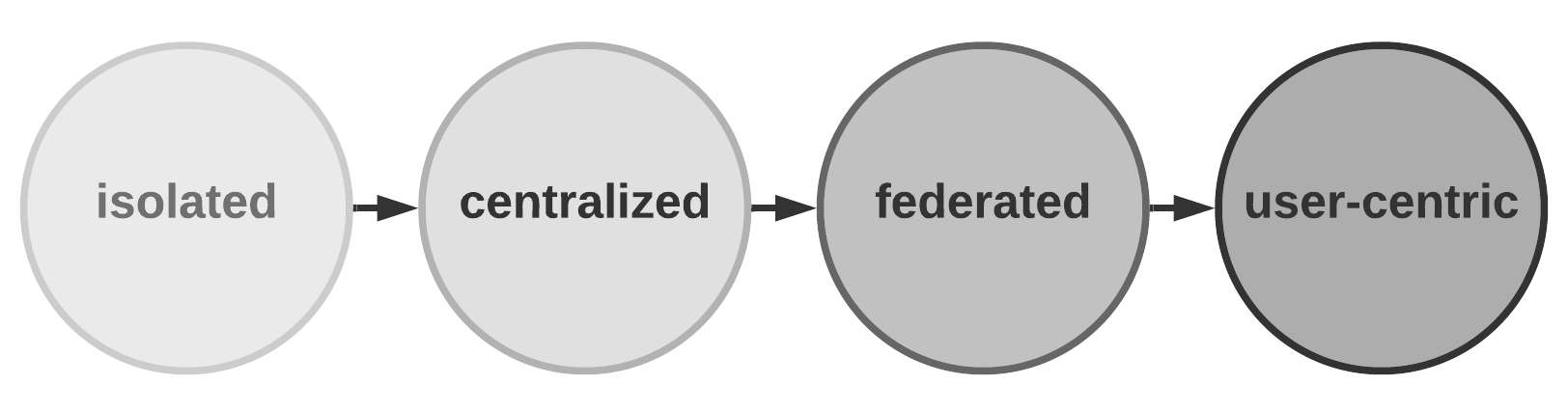}
\label{fig:model}}
\hspace{0.5cm}
\subfloat[Orthogonality of Identity Management Models]{
\includegraphics[width=.45\linewidth]{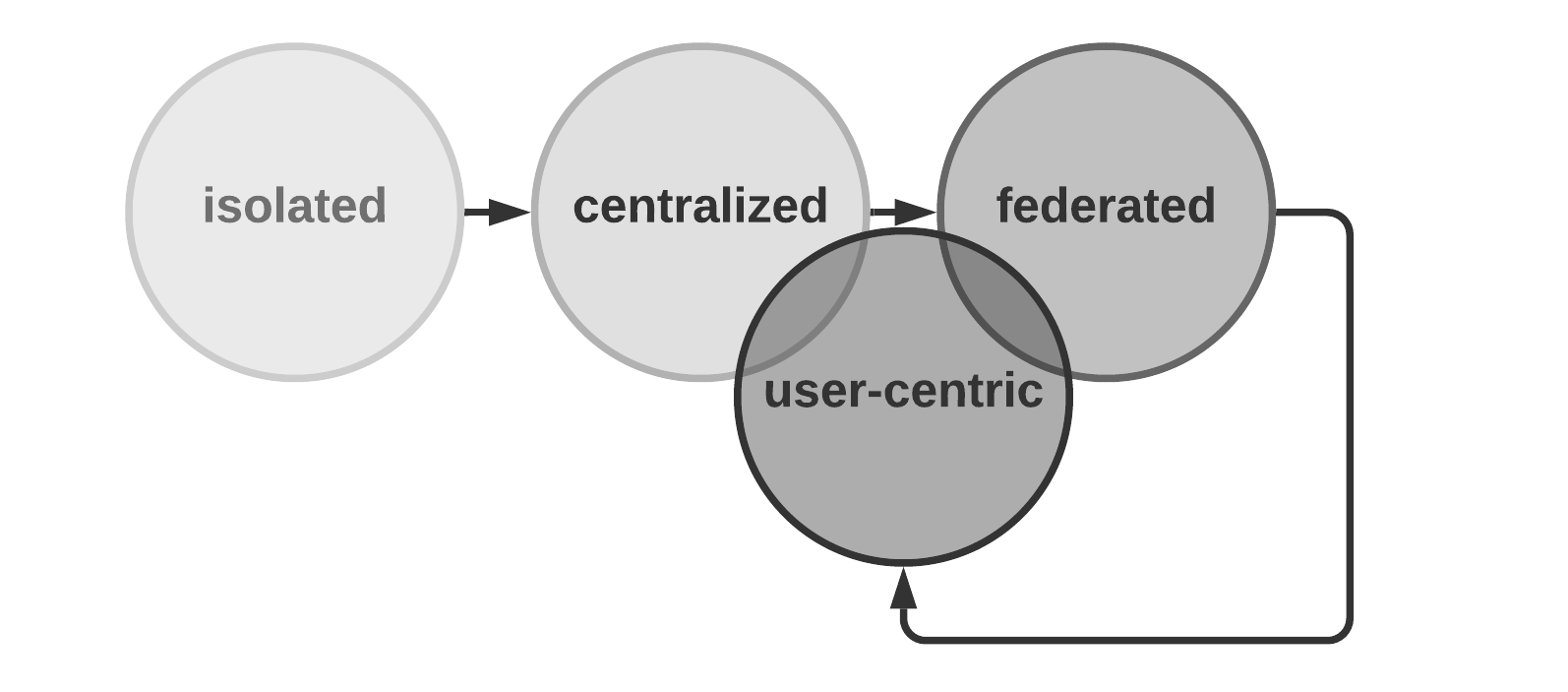}
\label{fig:ortho}}
\caption{Identity Management Models}
\end{figure}

The models are seen as evolution with almost no intersection, displayed in Figure~\ref{fig:model}. The models describe the topology and the source of truth, i.e., the user or another entity. Approaches can fit into two models at the same time, see Figure~\ref{fig:ortho}, e.g., if the IdM is user-centric but the SPs form a federation. In this case, SSI respectively UMA belong to two models. 

\subsection{The Identity Management Cube (IMC)}

In order to distinguish the approaches, we use the following dimensions.

\begin{description}
\item[Topology:] Topology of the IdM approach.
\item[Type of User:] Type of user, using the approach.
\item[Type of Service:] Type of service featured by the approach.
\end{description}

The \textit{topology} is orthogonal to user-centric and can be used as one category. Based on existing approaches, the topology can be described as follows. Isolated is left out of the category as it disappears due to the management overhead.

\begin{description}
\item[Centralized:] I\&AM per entity.
\item[With TTP:] I\&AM with several entities, where at least one TTP is involved. This applies to many cases of FIM and is, therefore, similar to federated.
\item[Without TTP:] I\&AM with several entities, where no TTP is involved. As it describes a distributed, completely decentralized structure, it addresses different approaches. Most cases of SSI belong to this category.
\end{description}

User-centric describes two things: a human user and user as source of truth. Other user types are computers, like servers, and IoT devices. Therefore, the second category is \textit{type of user}. The human user is further divided into user-centric and provider-centric, describing attribute handling.

\begin{description}
\item[User:] Divided into user-centric and provider-centric. This includes cases of UMA, SSI, but also SAML and OIDC.
\item[Computer:] Machine to machine (M2M) communication, for example. 
\item[IoT Device:] IoT devices usually have less computing power, which restricts computationally intensive cryptographic operations.
\end{description} 

Although an increasing number of web services are used, like Office 365, several services are non-web-based. In order to distinguish the \textit{type of service}, the following characteristics are set.

\begin{description}
\item[Non-Web Service:] M2M communication, but also local services.
\item[Background Web Service:] Services, which are need for interactive web services, like localization of the user's home organisation.
\item[Interactive Web Service:] Services the end-user uses.
\end{description}

\begin{figure}[h]
  \centering
    \includegraphics[width=0.45\textwidth]{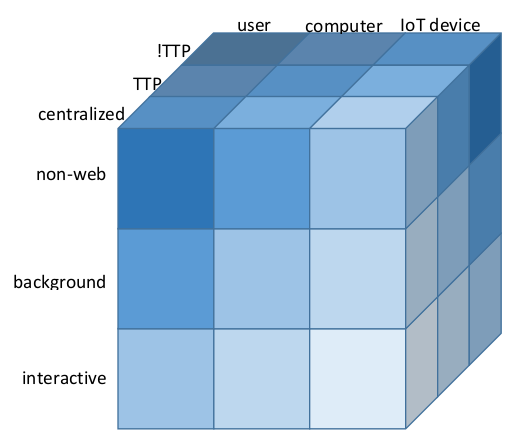}
    \caption{Identity Management Cube}\label{fig:cubic}
\end{figure}

As a result, the new model comprises three categories, topology, type of user, and type of service, displayed as axes. In reference to the Life cycle, Aspect, Layer (LAL) Brick~\cite{10.1007/978-3-540-30184-4_1}, the categories result in a cube. In Figure~\ref{fig:cubic}, the developed IdM cube including the labels of the different axes is shown. User-centric and provider-centric are thus left out for clarity reasons.

\subsection{IMC Applied to Current Approaches}

In order to depict the IMC, different IdM approaches are classified by the categories described above. As examples, centralized IdM with SAML federations in research and education (R\&E), OIDC in the web, UMA for private users, and SSI as new approach are chosen. In addition, IdM for servers, IoT, and with Active Directory (AD) are explained. 

\begin{figure}
\subfloat[IMC for SAML]{
\includegraphics[width=.45\textwidth]{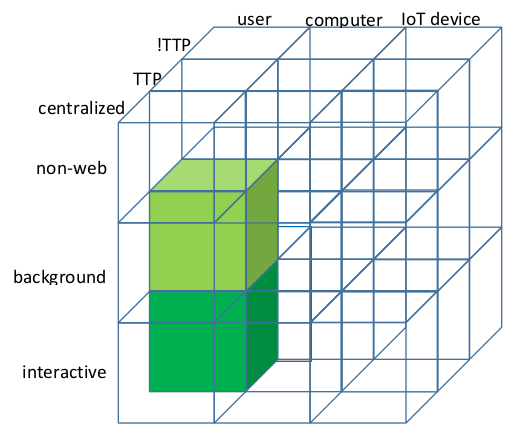}
\label{fig:cubic_saml}}
\hspace{0.5cm}
\subfloat[IMC for OIDC]{
\includegraphics[width=.45\linewidth]{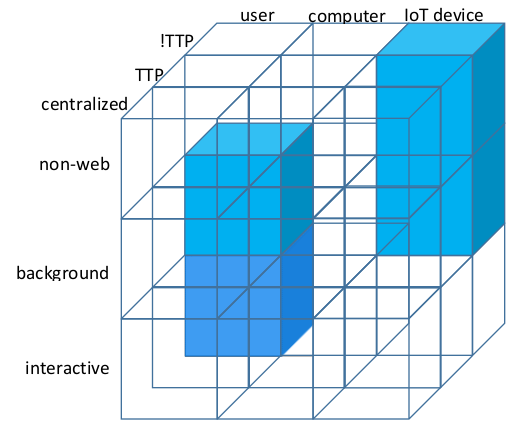}
\label{fig:cubic_oidc}}
\\
\subfloat[IMC for SSI]{
\includegraphics[width=.45\textwidth]{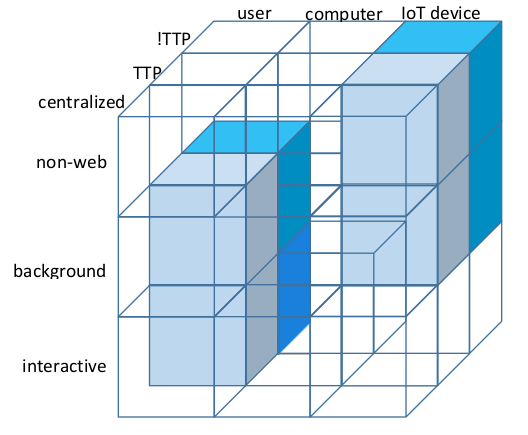}
\label{fig:cubic_ssi}}
\hspace{0.5cm}
\subfloat[IMC for Server]{
\includegraphics[width=.45\linewidth]{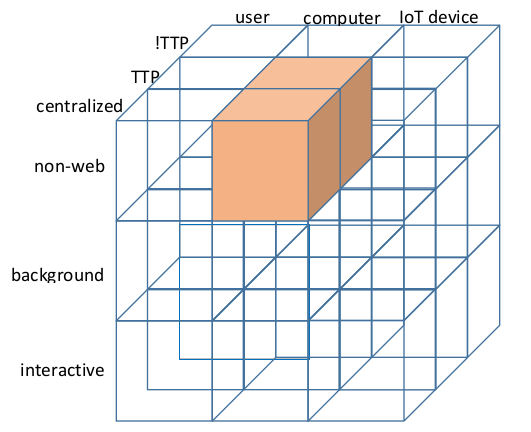}
\label{fig:cubic_server}}
\\
\subfloat[IMC for AD]{
 \includegraphics[width=0.45\textwidth]{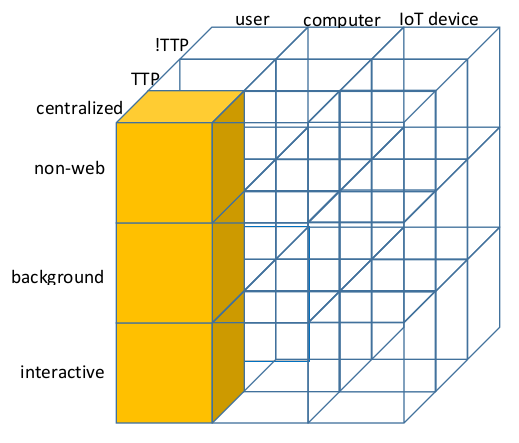}
\label{fig:cubic_ad}}
\hspace{0.5cm}
\subfloat[IMC for IoT]{
 \includegraphics[width=0.45\textwidth]{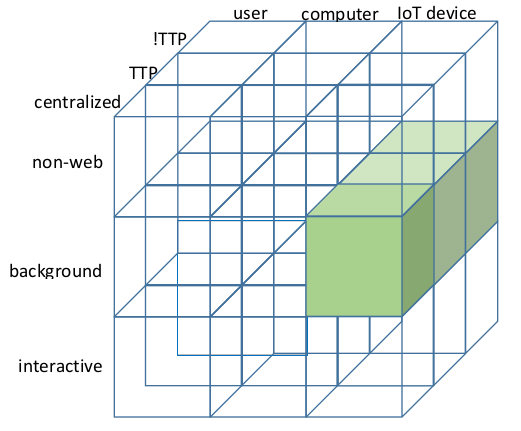}
\label{fig:cubic_iot}}
\caption{Identity Management Cubes Applied to Different Use Cases}
\end{figure}

\textit{SAML} is used in R\&E to let users access web services at research partners. It is based on lightweight directory access protocol (LDAP), databases, or even AD with the add-on federation. The entities form a federation, which relies on contracts with the federation operator. As a result, it has the following characteristics, as shown in Figure~\ref{fig:cubic_saml}.

\begin{description}
\item[Topology:] With federation tools as TTPs.
\item[Type of User:] User are humans, but the type is provider-centric.
\item[Type of Service:] Interactive web services for end users.
\end{description}
\textit{OIDC} is used in web as well, but is a more dynamic protocol without a TTP, based on OAuth. UMA is also developed on top of OAuth, but more user-centric. This can be seen in the characteristics, shown in Figure~\ref{fig:cubic_oidc}.
\begin{description}
\item[Topology:] Using Webfinger technology is without a TTP, but can be centralized in some use cases.
\item[Type of User:] Human end user in most cases, which can be either provider-centric (OIDC) or user-centric (UMA).
\item[Type of Service:] Typically interactive web for end users, but others types are also possible.
\end{description}
\textit{SSI} is seen as the new step in evolution of IdM, as the user is in control of everything. The concept is without a TTP, but it evolves to a topology with a TTP for scalability and performance reasons. Most approaches concentrate on interactive web services, though the concept could be applied to other services as well. SSI, therefore, has the following characteristics, displayed in Figure~\ref{fig:cubic_ssi}.
\begin{description}
\item[Topology:] Originally, SSI is without a TTP, but is evolving to centralized services.
\item[Type of User:] SSI focuses on the user, therefore, user-centric.
\item[Type of Service:] Interactive web services for end users.
\end{description}
Besides web application, \textit{servers} are run at the backend, which are normally access through keys. The public key is stored at the server, while the administrator is in possession of the private key. So, the service is non-web and it is typically either centralized or with a TTP. As a result, identity management for servers can be described as following, shown in Figure~\ref{fig:cubic_server}.
\begin{description}
\item[Topology:] Either centralized, with a centralized IdM, or with a TTP.
\item[Type of User:] Both, computer in M2M or human users are possible.
\item[Type of Service:] The services are typically non-web.
\end{description}
\textit{Centralized IdM with AD} is used in companies to enable employees to login at their computer, provision folders and shares, but also to access web services with single sign-on (SSO). It has the following characteristics, shown in Figure~\ref{fig:cubic_ad}.
\begin{description}
\item[Topology:] The AD itself is centralized.
\item[Type of User:] The human user is in focus, but the IdM is provider-centric. Additionally, Windows computer can be a user.
\item[Type of Service:] All types of services are possible, as it relies on Windows.
\end{description}
\textit{IoT devices} often communicate with Constrained Application Protocol (CoAP) instead of Hypertext Transfer Protocol (HTTP). The devices, which either lack a browser to perform user-agent based authorization or are input constrained, cannot make use of typical web protocols, like OAuth or SAML. One option is, e.g., to utilize shared keys, another is ACE-OAuth. ACE-OAuth maps OAuth methods to Authentication and Authorization for Constrained Environments (ACE). The characteristics are shown in Figure~\ref{fig:cubic_iot}.
\begin{description}
\item[Topology:] IoT networks are typically centralized managed, which can be with or without a TTP.
\item[Type of User:] The type is IoT device.
\item[Type of Service:] IoT devices are mainly background services.
\end{description}

\begin{figure}
\subfloat[Merged IMC]{
\includegraphics[width=.45\textwidth]{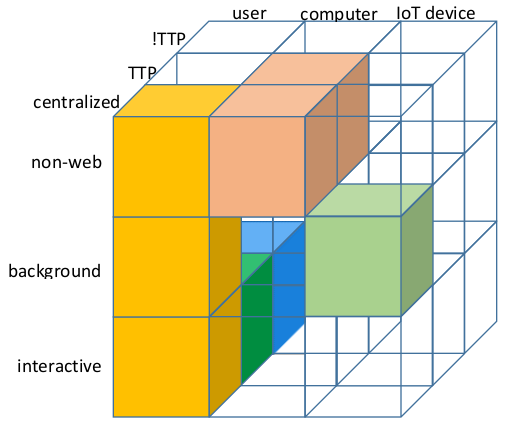}
\label{fig:cubic_applied}}
\hspace{0.5cm}
\subfloat[Merged IMC Including Unusual Use Cases]{
\includegraphics[width=.45\linewidth]{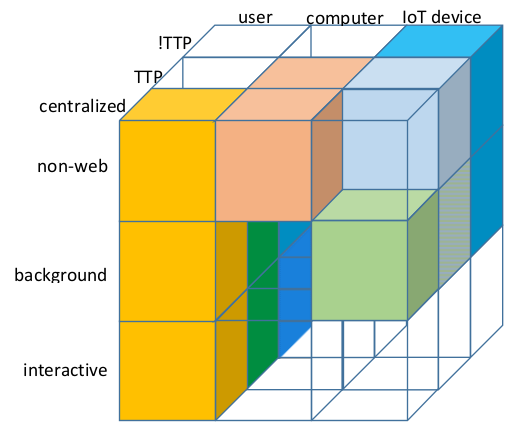}
\label{fig:cubic_max}}
\caption{Merged Identity Management Cubes}
\end{figure}

The selected approaches can be merged in one IMC. The colors are used as in the figure above: AD dark green, SAML yellow, OIDC dark blue, SSI light blue, servers orange, IoT light green. The cube illustrates that many approaches are used for interactive web and human users, while the protocols themselves could be used for other user cases as well. The figure at the same time visualizes the differences between the approaches. While AD is focused on centralized topology, SAML typically uses a TTP, while OIDC, UMA, and SSI tend to work without TTP. 
The most common type of service are used in Figure~\ref{fig:cubic_applied}, while Figure~\ref{fig:cubic_max} adds also unusual use cases.
Both figures show that the selected approaches do not cover all aspects of the IMC. SSI with a centralized party would partly fulfil the application of SAML, as it would double to with a TTP from the later figure. Especially these shared single cubes illustrate that interoperability between the approaches should be easily reached, while combining different approaches arranged in different cubes probably needs more effort and tools. Additionally, one can either have user-centric or service-centric. Most approaches cannot provide both, as trust into the attributes is missing.

\section{Morphology of Identity Management}\label{sec:addon}

In order to determine the degree of fulfilment, a uniform format is needed describing the approaches in more detail. Therefore, a characteristic similarly to a morphology is established. The morphology focusses on organizational aspects, while the IMC categorizes the technology. The authors of~\cite{iariajournal} describe the characteristics of Inter-FIM based on a morphology. As the characteristics need to comprise all IdM approaches and therefore relates to the IMC, the morphology is extended for the needs of universal IdM. In a next step, the morphology is mapped to the life-cycle, in order to clarify when which decision is taken. Last but not least, approaches are characterized by the morphology.

\subsection{Design of the Morphology}

The morphology describes the characteristics of the cooperation. \cite{iariajournal} uses cooperation structure, members, group structure, federation dimension, organizational dimension, duration, sort of collaboration, coordination, establishment, circle of trust, degree of commitment, and trust relationship. As this approach concentrates on Inter-FIM, the following characteristics can be left out or need changed.

\begin{description}
\item[Structure of Cooperation:] The structure described topology and cooperation customized for federations. The topology is described by the cube, while different aspects of the cooperation are part of the morphology.
\item[Cooperation:] Instead of FIM, Projects, and Communities, this characteristic is now described in ``Reason for Joining'' as well as ``Order''.
\item[Formalization] Differentiates between ``limited contract'' and ``cooperation agreements'', in order to describe the distinction of contracts.
\item[Dynamic of Joining:] Broader scope with ``stable'' and ``unstable''.
\end{description}

To describe different organizational aspects, other characteristics need to be added. Several characteristics relate to the IdM architecture.

\begin{description}
\item[Reason for Joining:] In order to differentiate between private usage and business reasons, this characteristic was included.
\item[Connectivity:] Describes the interaction between involved parties, which might have consequences for the architecture.
\item[Direction of Cooperation:] Broadens the scope.
\item[Administration:] Degree of automation, which relates to the architecture.
\item[Cooperation Structure:] Either ``hierarchical'' or ``heterarchical''.
\item[Level of Trust:] Trust between involved parties.
\item[Identities:] Included as it has implications for the architecture.
\end{description}

\subsection{Identity Management Morphology}

This results in a morphology, which includes more and broader characteristics. The morphology has the following categorize, as shown in Table~\ref{tab:morph}.

\begin{description}
\item[Initiation:] Initiation of the cooperation.
\item[Cooperation:] Settings of the cooperation.
\item[Coordination:] Settings of the coordination.
\item[Trust:] Trust between participating entities.
\item[Identities:] Settings of the identities.
\end{description}

\begin{table}
\caption{Morphology for Identity Management in Detail}
\label{tab:morph}
\begin{tabular}{p{2.8cm}||p{2.2cm}p{2.2cm}p{2.2cm}p{2.2cm}}
\multicolumn{5}{c}{Initiation}  \\ \hline
Reason for Joining & personal & social & economic & law  \\
Dynamic of Joining & \multicolumn{2}{c}{stable} & \multicolumn{2}{c}{unstable} \\ \hline
\multicolumn{5}{c}{Cooperation}  \\ \hline
Degree of Integration & autonomous & coordinated & integrated & \\
Connectivity &  \multicolumn{2}{c}{low} & \multicolumn{2}{c}{high} \\
Professional Limits &  user & R\&D & department & value chain  \\
Factual Limits & short & medium & long & permanent \\
Direction of Cooperation & vertical & horizontal & diagonal & \\
Order & strategy & project & R\&E & region \\
Locality &  local & regional & national & international \\
Organizational & micro & meso & macro & \\
Formalization & arrangement & limited contract & cooperation agreement & capital interweaving \\ \hline
\multicolumn{5}{c}{Coordination}  \\ \hline
Administration & manual & supported & automated & \\
Number of Participants & bilateral & simple & complex & \\
Group Structure & open & with limitations & closed & \\
Cooperation Structure  &  \multicolumn{2}{c}{hierarchical} & \multicolumn{2}{c}{heterarchical} \\
Sort of Coordination &  \multicolumn{2}{c}{implicit} & \multicolumn{2}{c}{explicit} \\ \hline
\multicolumn{5}{c}{Trust}  \\ \hline
Directness &  \multicolumn{2}{c}{ direct} & \multicolumn{2}{c}{transitive} \\
Circle of Trust & static & dynamic & virtual &  \\
Level of Trust & zero & low & medium & high \\  \hline
\multicolumn{5}{c}{Identities}  \\ \hline
Transparency & low &  medium & high & \\
Controllability & low & medium & high & \\
Identification & internal & external & combination & \\
Authentication Method & anonymous & simple &  2FA & MFA \\
Authentication Organization & internal & external & combination & \\
Authorization & internal & external & combination & \\
\end{tabular}

\end{table}

\textit{Initiation} comprises of reason for joining and dynamic of joining. The \textit{reason} can be ``personal'', ``social'', like in social media, ``by law'' or ``economic''. Economic reasons can further be split into ``time'', ``risk'', ``earnings'', ``competence'', ``costs'', ``pressure'', and ``protection''. Another distinction could be ``planned'', if necessary, and ``spontaneously event-driven''. The \textit{dynamic} is either ``stable'' or ``unstable'', i.e., it is either predictable or not.

The \textit{cooperation} itself is described by degree of integration, connectivity, professional limits, factual limits, direction of cooperation, order, locality, organizational, and formalization. Both, the degree of integration and the connectivity between partners, are part of the networking between partners. The \textit{degree of integration} can either be ``autonomous'', ``coordinated'' or ``integrated''. This means that either the partners work autonomous, coordinated towards a goal. Integrated can be a fusion of organizations. The \textit{connectivity} has two steps: ``low'' and ``high''. It partly relates to integration. The next category are both limits, professional and factual. \textit{Professional limits} describe which organization part is involved in the cooperation. It can be ``research'', a ``department'', the complete ``value chain'', or just one or more ``users''. \textit{Factual limits} are described by ``permanent'' or ``restricted''. Restricted can further be split into ``short'', ``medium'', and ``long''. The \textit{direction of cooperation} depicts how close both economic levels are related. ``Horizontal cooperation'' describes the cooperation of companies of the same business or same level of the value chain, while ``vertical cooperation'' is a cooperation between organizations of different economic levels, like retail company and production company. A cooperation is ``diagonal'', if all involved companies are neither on the same economic level nor business, e.g., travel company and food company. The \textit{order} characterizes the reason for the cooperation, which is ``strategic'', a ``project'', ``R\&E'', or based on the ``region''. Both, the locality and the organizational are dimensions of the cooperation. The \textit{locality} of the cooperation is either ``local'', ``regional'', ``national'', or ``international''. A national federation are the R\&E federations, like SWITCHaai in Switzerland. eduGAIN is the international umbrella federation for the national pendants.

The \textit{organizational} dimension describes the viewing plane of the cooperation. Terminology from economics is used. ``Micro'' plane consists of one single entity, while the ``meso'' plane comprises of several organizations, e.g., in a federation. The ``macro'' plane shows the cooperation of cooperation, e.g., an inter-federation. The \textit{formalization} classifies the kind of formality between the entities. While an ``arrangement'' can be oral or somehow written, a ``contract'' is divided into limited length and cooperation. The last step is a ``capital'' interweaving of the involved entities. An example for an arrangement is the usage of social media for end users, while contracts are typical for projects. The formalization also describes the binding intensity, which is the degree by with the involved entities give up their autonomy.

The \textit{coordination} explains the management of the cooperation, which consists of administration, number of participants, group structure, order, and sort of coordination. The number of participants is related to the group structure. Open cooperation do not have a firm number of participants. Closed cooperation allow simple as well as bilateral structures. The coordination further relates to trust. The \textit{administration} can be ``manual'', ``supported'' or ``automated''. The \textit{number of participants} is strongly related to cooperation. The participating entities can either have a ``bilateral'' agreement, the cooperation can have a ``simple'' structure, or it can be ``complex''. While bilateral cooperation still work with duplicated user bases, this is not possible with more entities involved. Simple networks can be realized with security assertion markup language (SAML) federations, while complex structures are also more complex for technical realization. OpenID Connect (OIDC) can be used for it. The \textit{group structure} is either ``open'', ``with limitations'' or ``closed''. OIDC is typically open, while SAML federations have limitations in R\&E or are closed in industry. The \textit{cooperation structure} is ``hierarchical'' or ``heterarchical'', when all partners are more or less of the same level. The \textit{sort of coordination} has two possible values: ``implicit'' or ``explicit''. With an explicit coordination, the integration of an institutional coordination instance is supported. An implicit coordination needs a local coordination.

\textit{Trust} between entities is the result of several different factors, like recommendation or past experience. Within the morphology, only the basics for the cooperation are described, which includes directness, dynamics, and the average level of trust. The circle of trust (CoT) relates to the sort of cooperation. If the group structure is limited, then the CoT can be static. If the number of participants is complex, is the CoT virtual as not all the information about all participants cannot be fully known. Direct trust implicates static or dynamic CoT. \textit{Directness} describes how the trust between two entities is derived. The trust is either ``direct'' or ``transitive / indirect'', via another entity. The \textit{dynamics} characterize the trust over time, which is either ``static'' or ``dynamic''. Last but not least, the \textit{level of trust} can be ``zero'', ``low'', ``medium'', or ``high''.

As final category, \textit{identities} are classified. Identities especially describe factors of trust and also user-centric features. This includes transparency, controllability, identification, authentication, consisting of methods and organizational factors, as well as authorization. The \textit{transparency} is either ``low'', ``medium'' or ``high''. The same characteristics can be applied to \textit{controllability}. The \textit{identification}, \textit{authentication}, and \textit{authorization} can be done ``internally'', ``externally'', or in a ``combination'' of different entities. The \textit{authentication methods} describe the sort of credentials used, which is either ``anonymous'', ``simple'' (like a password), ``second-factor authentication (2FA)'' or ``multi-factor authentication (MFA)''. In order to reduce the complexity of the morphology, suited characteristics can be left out. In the next step, special characteristics for the use case, like topology of federation can be added. This depends on the specific use case.

\subsection{Morphology Mapped to Life Cycle}

The morphology can be mapped to the life cycle of IdM, helping starting cooperation to identify their framework. The life cycle is similar to the Deming Cycle~\cite{deming}, which has the phases plan, do, check, act. The Deming Cycle is an iterative four-step management method used in business to control improvements of processes and products. It can be applied to service management, security management, and many other, like identity management. The life cycle of IdM has the phases initiation, agreement, cooperation, reconsideration and improvement, and termination. Reconsideration can either lead to improvement or termination. The phases of the \textit{IdM life cycle} have the following characteristics.

\begin{description}
\item[Initiation:] A purpose leads to the initiation of the cooperation.
\item[Agreement:] After discussions, an agreement is signed, describing the framework of the cooperation. IdM should be a part of the agreement. Otherwise, the parties need to agree on IT aspects outside of the agreement.
\item[Cooperation:] The cooperation is starting. In many cases, the cooperation is starting slowly, setting everything in place. Then there is a hype of cooperation, where everything is running and the original purpose is hopefully met. In IT, the start requires work, setting up the infrastructure.
\item[Reconsideration:] It describes if the cooperation is proceeded and if changes need to be made. The same appears for IdM.
\item[Improvement:] The changes lead to improvements, which are implemented.
\item[Termination:] If the purpose is met or other reasons lead to the end of the cooperation, the IdM is also terminated for the project.
\end{description}

\begin{figure}
\subfloat[Morphology mapped to Identity Management System Life Cycle]{
\includegraphics[width=.45\textwidth]{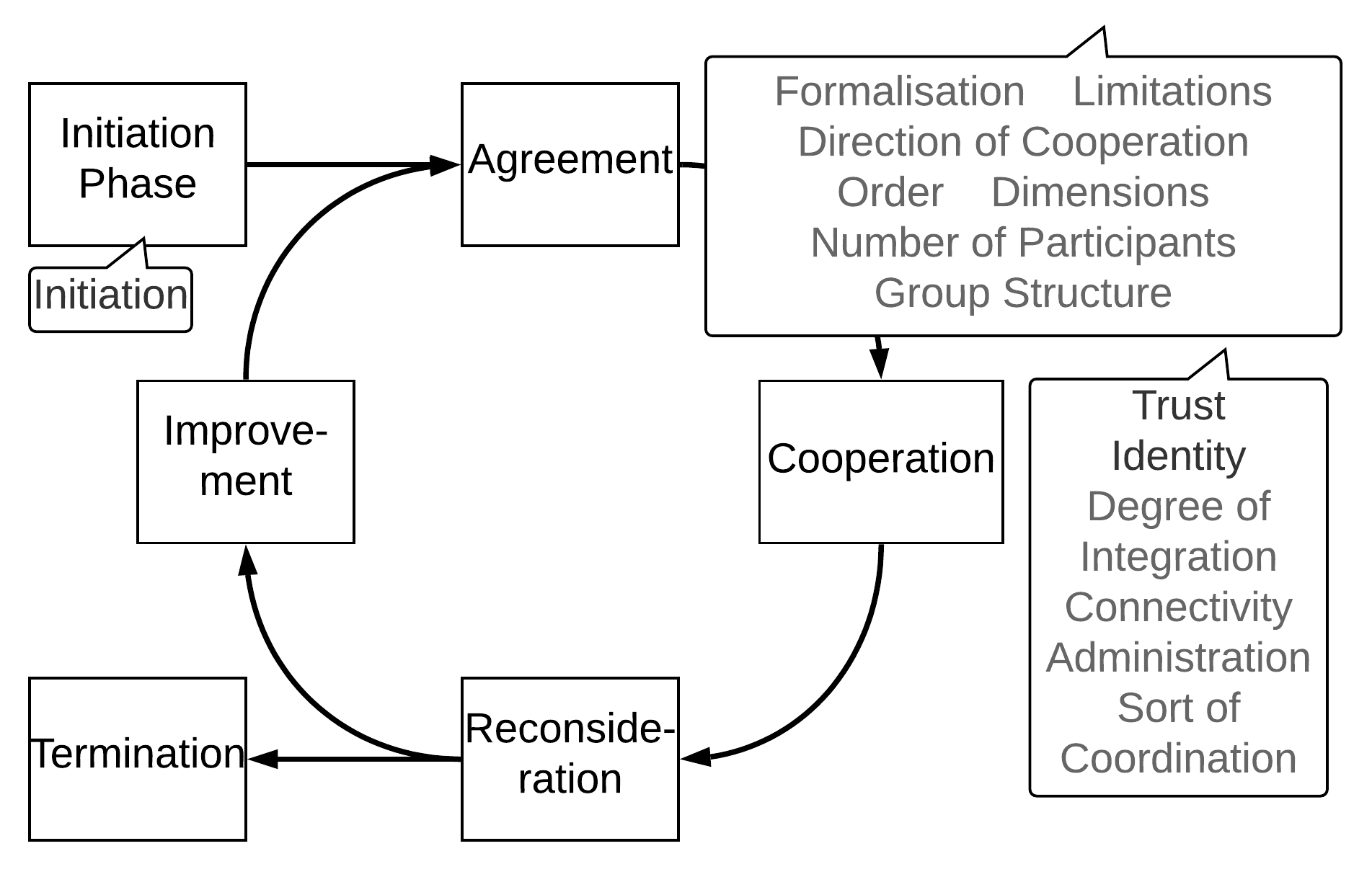}
\label{fig:idmlifecycle}}
\hspace{0.5cm}
\subfloat[Morphology mapped to User Life Cycle]{
\includegraphics[width=.45\linewidth]{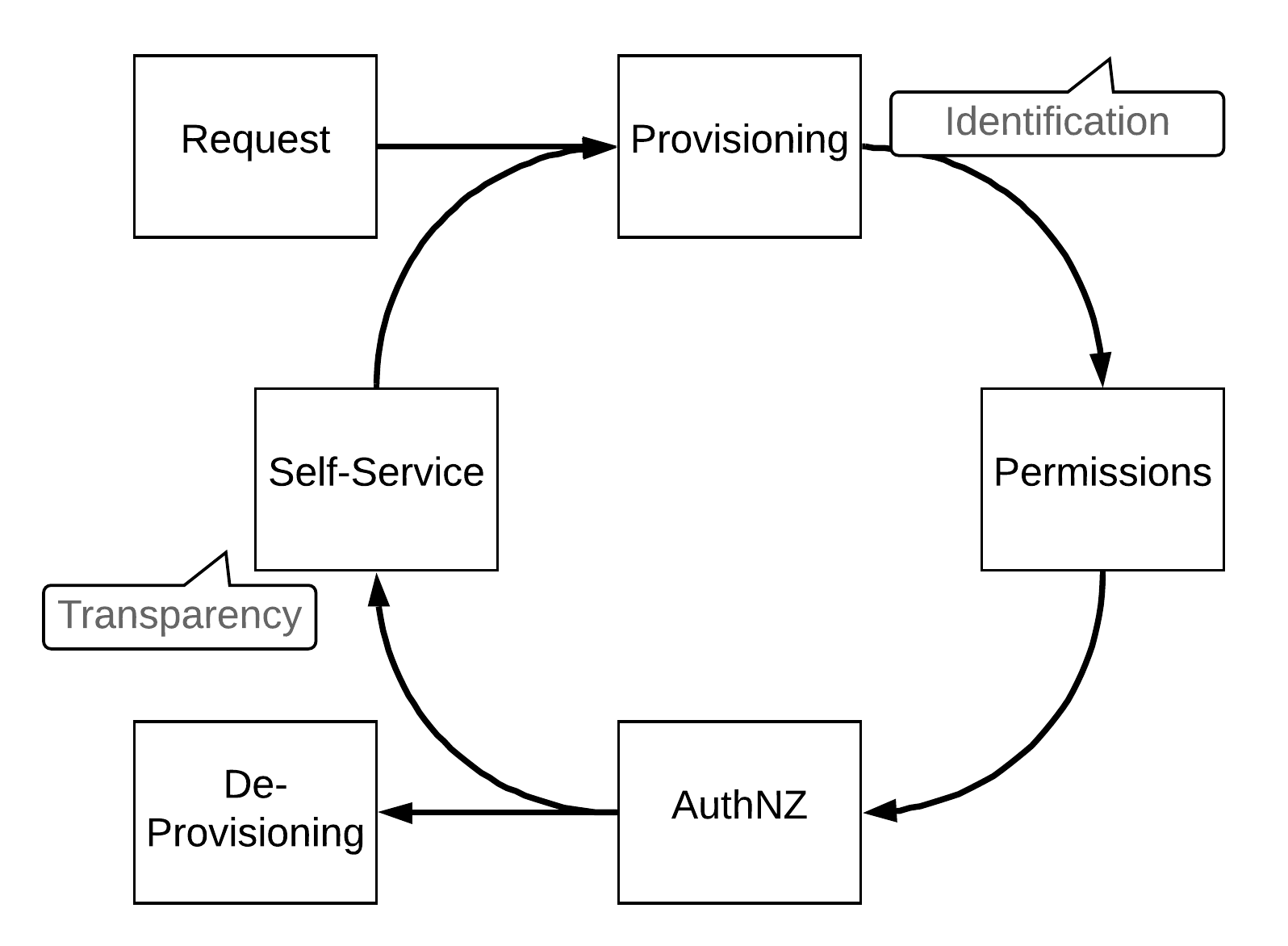}
\label{fig:userlifecycle}}
\caption{Morphology mapped to Life Cycles}
\end{figure}

The morphology, described in the previous section, can be mapped to the life cycle. This helps to gain a better picture of the required decisions, as shown in Figure~\ref{fig:idmlifecycle}. The \textit{initiation} phase comprises both characteristics of the morphology of initiation, which means ``Reason of Joining'' and ``Dynamics of Joining''. \textit{Cooperation} and \textit{coordination}, have characteristics in ``Agreement'' and ``Cooperation''. This is the case as some characteristics are decided at the agreement, while others have more impact on the cooperation. The agreement thereby features: ``Formalization'', ``Limitations'', ``Direction of Cooperation'', ``Cooperation Structure'', ``Dimensions'', ``Number of Participants'', and ``Group Structure''. The cooperation as a result includes ``Trust'', ``Identity'', ``Degree of Integration'', ``Connectivity'', ``Administration'', and ``Sort of Cooperation''. During \textit{reconsideration} every aspect is re-evaluated. Some aspects are enhanced during \textit{improvement}. The life cycle is terminated, if the cooperation ends. 

The IdM life cycle includes the user life cycle, because users change throughout a project or life cycle. During a project, users leave, while others join. This is also the case for IdM in organizations. In the end, every user account needs to be closed. The \textit{life cycle of the user} includes the following phases.

\begin{description}
\item[Request:] The user requests an account at an IdMS.
\item[Provisioning:] The account is provisioned (attributes, roles, and permissions).
\item[Identification:] The user identifies himself.
\item[Authentication and Authorization:] First authentication, then authorization.
\item[Self-Service:] The user can access the self-service.
\item[De-Provisioning:] In the end, the user account is de-provisioned.
\end{description}

The life cycle of the user has only few interactions with the morphology: trust into the service provider during \textit{request}, then identification, authentication, authorization themselves as well as transparency in the \textit{self-service} phase. This is also visualized in Figure~\ref{fig:userlifecycle}.

As a result, IdM models describe the approaches in general, while the morphology details aspects of the cooperation. The mapping of morphology with the life cycles explains the order of the actions, which need to be taken. This can guide projects and organizations to identity management processes. Nevertheless, a decision matrix for choosing the best fitting approach is missing, although the cube gives a first hint. Additionally, interfaces to already established processes, like service management and security management, are needed.

\subsection{Morphology Applied to Current Approaches}

Following, the morphology is applied to centralized IdM with AD and the differences for SAML, OIDC, SSI, server, and IoT devices are described. \textit{Centralized IdM with AD} is typically used in companies. Depending on whether they have cooperation, several branches or not, the complexity is different. Also depending on the point of view, e.g., user or company, different properties can be coloured. Let us assume the company in this example just uses AD for its users, while they have other methods for cooperation. Therefore, the following morphology can be formed, see Table~\ref{tab:admorph}. The reason for joining is economic, while the dynamic is stable. AD was introduced at some point in time. When regarding the cooperation, the cooperation within the company is considered. The integration is therefore integrated. The connectivity is high as all participants work together. AD was introduced for the complete value chain of the company. It should be a permanent solution, though technologies and decisions change. The direction of cooperation cannot be described by the categorization. The order was based on a strategy, while the company is local. The organizational factor is micro. The cooperation within the company depends on contracts with its employees. The administration is hopefully supported, while the number of participants can be described with bilateral. The group structure is at least currently closed, while the coordination is hierarchical and explicit. The trust is direct, rather static, with medium trust, as all employees needed to submit papers. The identities are managed within the company, with simple and second factors.

\begin{table}[!htb]
\caption{Morphology for Centralized Identity Management with AD}
\label{tab:admorph}
\begin{tabular}{p{2.8cm}|p{2.2cm}p{2.2cm}p{2.2cm}p{2.2cm}}
\multicolumn{5}{c}{Initiation}  \\ \hline
Reason for Joining & personal & social & \cellcolor{gray}economic & law  \\
Dynamic of Joining & \multicolumn{2}{c}{\cellcolor{gray}stable} & \multicolumn{2}{c}{unstable} \\ \hline
\multicolumn{5}{c}{Cooperation}  \\ \hline
Degree of Integration & autonomous & coordinated & \cellcolor{gray}integrated & \\
Connectivity &  \multicolumn{2}{c}{low} & \multicolumn{2}{c}{\cellcolor{gray}high} \\
Professional Limits & user & R\&D & department & \cellcolor{gray}value chain  \\
Factual Limits & \cellcolor{gray}short & \cellcolor{gray}medium & \cellcolor{gray}long & \cellcolor{gray}permanent \\
Direction of Cooperation & vertical & horizontal & diagonal & \\
Order & \cellcolor{gray}strategy & project & R\&E & region \\
Locality & \cellcolor{gray}local & regional & national & international \\
Organizational & \cellcolor{gray}micro & meso & macro & \\
Formalization & arrangement & \cellcolor{gray}limited contract & cooperation agreement & capital interweaving \\ \hline
\multicolumn{5}{c}{Coordination}  \\ \hline
Administration & manual & \cellcolor{gray}supported & automated & \\
Number of Participants & \cellcolor{gray}bilateral & simple & complex & \\
Group Structure & open & with limitations & \cellcolor{gray}closed &  \\
Cooperation Structure  & \multicolumn{2}{c}{\cellcolor{gray}hierarchical} & \multicolumn{2}{c}{heterarchical} \\
Sort of Coordination &  \multicolumn{2}{c}{implicit} & \multicolumn{2}{c}{\cellcolor{gray}explicit} \\ \hline
\multicolumn{5}{c}{Trust}  \\ \hline
Directness &  \multicolumn{2}{c}{\cellcolor{gray}direct} & \multicolumn{2}{c}{transitive} \\
Circle of Trust & \cellcolor{gray}static & dynamic & virtual &   \\
Level of Trust & zero & low & \cellcolor{gray}medium & high \\  \hline
\multicolumn{5}{c}{Identities}  \\ \hline
Transparency & low & \cellcolor{gray}medium & high \\
Controllability & low & \cellcolor{gray}medium & high \\
Identification & \cellcolor{gray}internal & external & combination & \\
Authentication Method & anonymous & \cellcolor{gray}simple & \cellcolor{gray}2FA & MFA \\
Authentication Organization & \cellcolor{gray}internal & external & combination & \\
Authorization & \cellcolor{gray}internal & external & combination & \\
\end{tabular}
\end{table}

The entities of \textit{SAML} in R\&E form federations, which are spread over regions, countries, and the world. The entities have contracts with federation operators, which have contracts with the inter-federation operator. The coordination between the entities is rather low. As the identities are managed by the home organization, trust is lower. Therefore, the following morphology can be seen.

\begin{description}
\item[Initiation:] Individuals join for R\&E, while companies have economic reasons. The dynamic is rather stable, as entities have to sign on contracts.
\item[Cooperation:] The cooperation is autonomous, only little coordinated by the federation and inter-federation operators. The connectivity between the entities is low, as there are many different services within a federation and only a small percentage of users will use the specific service of a service provider. Mostly, the entities have contact with the federation operators. The cooperation is limited to research, while the time depends on several reasons. The cooperation can be vertical as well as horizontal. The locality is national or international in most cases. Also regional federations are established. The organization form is either meso or macro, depending on the type of federation. Federations are formalized by contracts with the federation operator and partly arrangements between entities.
\item[Coordination:] The administration is supported with manual steps needed. As contracts need to be signed, the number of participants is simple and the group structure is with limitations. The order is more heterarchical than hierarchical, while the coordination is more implicit than explicit.
\item[Trust:] Trust is transitive via federation or inter-federation operator. With a static number of participants, the circle of trust is also static with little dynamics. The level of trust is low or medium, depending on separate means.
\item[Identities:] Since communities with additional attribute authorities were formed and other means of identification are in use, authorization and identification are either internal or a combination, while authentication is internal. Transparency and controllability are rather low as a result of the structure.
\end{description}

IdM with \textit{OIDC} distinguishes from SAML as the protocol is dynamic and the widely known use case is web authentication. For OIDC, the initiation can have several reasons, therefore, the dynamic is unstable. The cooperation is loose, which is true for the coordination as well. Trust is rather low, but can be stepped up with a second factor. The different constellations also have impact on the identities.
\textit{SSI} is different as the user is in control of the attributes, which then impacts trust and identities. If a company hosts their \textit{servers} in-house, then the cooperation is within the company and maybe with other offices. \textit{IoT devices} can be used at home as well as at organizations. The trust into the devices is typically low, as others might manipulate the device without notice.

\section{Discussion}\label{sec:five}

We characterized IdM approaches in two ways: the IMC describes the technical aspects, followed the morphology for organizational aspects. In order to compile an overview of IdM approaches, we noticed intersections between existing IdM models. These intersections helped us to identify categories, which are needed to differentiate IdM approaches. The three categories topology, type of user, and type of service are arranged in a cube, the IMC. IMC clarifies the characteristics type of user and topology. Additionally, the perspective is made clear, i.e., user-centric or provider-centric. While an approach could belong to two models used beforehand, it can be clearly classified by the IMC. With the flexibility of the three categories in mind, future approaches should be able to be characterized. In the next step, we applied different IdM approaches to the cube. These approaches were typical web services for end users, but also servers and IoT, resulting in a colourful IMC. Some use cases are more typical than others. Besides this fact, the application was straight forward and showed us similarities and differences between the approaches. These findings might indicate a possible combination of approaches. It is noticeable that trust and user-centric are not featured together in the shown examples. The IMC can, therefore, help to combine different IdM approaches and explore missing tools.

In a next step, a morphology for IdM was developed. The morphology describes different aspects of a cooperation. In this case, the categories initiation, cooperation, coordination, trust, and identities with their categorizations need to be regarded. The relationship between morphology and the different life cycles were shown in a next step. For guidance, the morphology can be used to speed up the implementation and evaluation in a later step. The morphology was then mapped to different approaches. A certain variance is seen, which depends on the actual implementation. Nevertheless, organizational settings are made clear. This does not include internal processes, which will be regarded in future work.

Both, the IMC and the morphology, do not describe IdM in all aspects, but help to categorize different approaches, use cases, and their implementation. The categorization helps within the life cycle of IdM to mix different approaches, see missing tools, and to regard all relevant aspects.

\section{Conclusion and Future Work}\label{sec:summary}

Identities are everywhere nowadays. With the growing number of internet users and accounts, more servers are used. With new opportunities, also new use cases come into sight. The IdM model was developed before the hype of blockchain. New approaches were established since then.

In this paper, we introduced a broad classification of IdM. The existing IdM modes were extended to an IMC with three axes. Topology, type of user, and type of environment describe the IdM approach in more details while still being vague about the actual protocols. The IMC was applied to different approaches, showing silos as well as approaches, which should be comparably easy to interoperate with additional tools. This showed that many aspects rely on the actual implementation within the organization. Also, it visualizes a trade-off between user control and trust into attributes. The IMC was extended by a morphology of IdM, which describes the characteristics of cooperation. This morphology was mapped to the life cycle of users and IdM in a further step. The result of the mapping can help to distinguish relevant questions during a cooperation. Both methods, the IMC and the IdM morphology, combined provide a comprehensive characterization of IdM approaches. This helps to choose suitable approaches for an organization or cooperation. Furthermore, needed tools for interoperability can be explored. An integration into processes and a guide to choose the best fitting IdM approach were left out and will be further work. The methods also reveal that interesting features for a holistic IdM have not been designed yet.

In order to create one holistic IdM framework, integrating different IdM approaches, an architecture is being developed. This architecture is extended by service models, visualizing needed processes. As a another step, processes interacting with already established management processes are investigated. To decide for the best fitting IdM approach, a decision matrix is created, all helping to ease the use and improve the quality of IdM.

\bibliographystyle{splncs04}
\bibliography{classification}

\end{document}